\documentclass{emulateapj}

\usepackage{amsmath,amssymb}
\usepackage{apjfonts}

\submitted{Submitted to ApJ}

\shortcites{astro-ph/0302213,astro-ph/0302209,2001ApJ...561..521A,Spergel:2006hy,Dijkstra:2003vg}
\def\gsim{\;\rlap{\lower 2.5pt
 \hbox{$\sim$}}\raise 1.5pt\hbox{$>$}\;}
\def\lsim{\;\rlap{\lower 2.5pt
   \hbox{$\sim$}}\raise 1.5pt\hbox{$<$}\;}

\newcommand{\msun}{\ensuremath{{\rm M}_\sun}}
\newcommand{\Mpc}{\ensuremath{\mathrm{Mpc}}}
\newcommand{\Tvir}{\ensuremath{T_\mathrm{vir}}}
\newcommand{\Tvirmin}{\ensuremath{T_\mathrm{vir,min}}}
\newcommand{\klvn}{\ensuremath{\mathrm{K}}}

\shorttitle{Clustering and Feedback During Reionization}
\shortauthors{Kramer, Haiman, and Oh}

\begin{document}

\title{Feedback from Clustered Sources During Reionization} 
\author{Roban Hultman Kramer, Zolt\'an Haiman}
\affil{Department of Astronomy, Columbia University, 550 West 120th Street, New York, NY 10027l} 
\author{S. Peng Oh}
\affil{Department of Physics; University
of California; Santa Barbara, CA 93106}
\email{roban@astro.columbia.edu, zoltan@astro.columbia.edu, peng@physics.ucsb.edu} 

\begin{abstract}
The reionization history of the intergalactic medium (IGM) at high
redshift ($z\gsim 6$) was likely strongly shaped by several global
feedback processes.  Because the earliest ionizing sources formed at
the locations of the rare density peaks, their spatial distribution
was strongly clustered.  Here we demonstrate that this clustering
significantly boosts the impact of feedback processes operating at
high redshift. We build a semi-analytical model to include feedback
and clustering simultaneously, and apply this model to the suppression
of star--formation in minihalos due to photoionization. The model is
built on the excursion-set-based formalism of
\citet*{astro-ph/0403697}, which incorporates the clustering of
ionizing sources, and which we here extend to include suppression of
star formation in minihalos.  We find that clustering increases the
mean HII bubble size by a factor of several, and it dramatically
increases the fraction of minihalos that are suppressed, by a factor
of up to $\sim 60$ relative to a randomly distributed population. This
enhanced suppression can significantly reduce the electron scattering
optical depth $\tau$, as required by the three--year data from the
{\it Wilkinson Microwave Anisotropy Probe (WMAP)}. We argue that
source clustering is likely to similarly boost the importance of a
variety of other feedback mechanisms.
\end{abstract}

\section{Introduction}

The three-year data on the polarization of the cosmic microwave
background from the {\it Wilkinson Microwave Anisotropy Probe (WMAP)}
satellite have provided an important new constraint on the
reionization history of the intergalactic medium (IGM). The {\it WMAP}
experiment has measured the optical depth $\tau$ to free--electron
scattering between us and the recombination epoch at $z\approx1100$.
The result, $\tau=0.09\pm0.03$, implies a redshift for instantaneous
reionization of $z\approx11$ \citep{Spergel:2006hy}, and can be used
to rule out a tail of significant ionization extending out to higher
redshifts.  Another constraint on reionization follows from the
Gunn--Peterson (GP) troughs detected in the spectra of bright quasars
at $z\approx6$. These GP troughs probe the neutral hydrogen along the
line of sight, indicating that the last $\sim0.1\%$ of the neutral
fraction was rapidly disappearing around this epoch
\citep{astro-ph/0602375, MH04}.

The relatively short reionization history implied by the combination
of these constraints requires a strong suppression of ionizing
radiation from ``minihalos'' with virial temperatures $\Tvir < 10^4\
\klvn$ \citep{Haiman:2006si}.  A variety of feedback mechanisms might
play a role in this suppression \citep[e.g.][]{HH03}.  In this paper,
we will explore one such feedback mechanism -- the suppression of gas
infall and star formation in minihalos in photoionized regions -- and
focus on how the clustering of the ionizing sources enhances this
effect.

Many other feedback mechanisms may also play a role in driving the
history of reionization. Some (such as the buildup of an X-ray
background) may affect the whole IGM more or less uniformly and
simultaneously, but most will be localized to within a small fraction
of the Hubble distance, at least for some time. Examples of such
localized feedback mechanisms are (1) the first generations of stars
will enrich the intergalactic medium with metals
\citep{2001ApJ...555...92M,2001ApJ...561..521A}, which will affect the
formation and evolution of subsequent generations of stars
\citep{2000ApJ...528L..65T,2001ApJ...552..464B,2002A&A...382...28S,2003ApJ...588L..69W};
(2) ionizing UV photons can both catalyze the formation of
\citep{RGS2002b, OH02, OShea05}, and directly destroy
\citep{1997ApJ...476..458H} $\mathrm{H}_2$ molecules, which will
affect the ability of gas to cool and form structures
\citep{1997ApJ...484..985H,Haiman:1999mn,2000ApJ...533..594C}, (3) gas
in regions that have been ionized may not be able to settle into the
shallow potential wells of small dark matter halos, exacerbating the
${\rm H_2}$--photodissociating effects of a Lyman Werner background
\citep{1997ApJ...484..985H, MBH06} and increasing the minimum size
required for a halo to contain collapsed baryonic structures
\citep{RGS2002a,RGS2002b,OH03}.

Since these mechanisms operate over a limited length scale, their
effects will depend strongly on the spatial distribution of halos
hosting ionizing sources. Numerical simulations are a promising way to
address feedback among clustered sources, since they capture the full,
three-dimensional relationships among the host halos.  However, the
dynamic range required to resolve the small minihalos, within a large
enough cosmic volume to be representative, remains a challenge,
especially in simulations that include radiative transfer \citep[see,
e.g.][]{astro-ph/0603199}.  Semi-analytical models avoid the problems
associated with the large dynamical range; they are also an efficient
way to explore parameter space and serve as important sanity checks
for more complicated simulations.

We believe that this paper describes the first semi--analytical model
to include feedback and clustering simultaneously as a way to suppress
ionizing radiation from minihalos.  Prior work on the clustering of
halos has focused on the problem of deriving analytical expressions
for the bias between the distribution of dark matter halos and of mass
\citep{astro-ph/0205276}; \citep[see
also][]{astro-ph/9512127,astro-ph/0410088}. These results have been
applied to the distribution of minihalos at high redshifts:
\citet{2001ApJ...551..599H} included the bias to account for the
effects of minihalo photoevaporation; \citet{astro-ph/0411035}
likewise characterized the effect of minihalo clustering on the
propagation of ionization fronts through the IGM. Minihalos were
considered to be sinks of ionizing photons in both of these works
\citep[as well as in][]{astro-ph/0511623,astro-ph/0505065}. In
contrast, in the present paper, we examine minihalos as potential
sources of ionizing radiation.  This means that the radius of
influence of our feedback mechanism cannot be simply evaluated at a
fixed time-step; instead, it is a function of the entire ionization
history of the region, including the history of mergers between
ionized regions. The preceding methods cannot be easily be adapted to
such a calculation, and we have been driven to employ a novel
Monte-Carlo means of simulating the coupled history of ionization and
feedback around any given point. Our primary conclusion is that,
because of their highly clustered distribution, local feedback effects
can dramatically suppress star formation in minihalos.

The rest of this paper is organized as follows. In
\S~\ref{section:model}, we describe our model for reionization,
extending the formalism of \citet*{astro-ph/0403697} to include an
approximate treatment of biased halo distribution and feedback.  In
\S~\ref{section:results}, we report our results for the reionization
history, and discuss the impact of clustering on the overall
suppression of minihalos, on the size distribution of ionized bubbles,
and on the optical depth to electron scattering. In
\S~\ref{section:discussion}, we discuss several limitations and
possible extensions and future applications of our model. Finally, in
\S~\ref{section:conclusions}, we summarize our conclusions and the
implications of this work.

\section{Modeling Reionization With Clustering and Feedback}
\label{section:model}

We have chosen to implement the suppression of minihalos in ionized
regions \citep{1996ApJ...465..608T,Dijkstra:2003vg,OH03}, both to
understand the effects of this particular mechanism, and as an example
of localized feedback mechanisms in general. We parameterize the
suppression effect by setting different minimum virial temperatures
for a halo to contribute to reionization $\Tvirmin{}$, in ionized and
neutral regions:
\begin{equation}
\Tvirmin{} = \left\{
\begin{aligned} 
10^2\ \klvn &\text{ in neutral regions,}\\
10^4\ \klvn &\text{ in ionized regions.}
\end{aligned}
\right.
\end{equation}
We will hereafter refer to halos with virial temperatures between
$10^2\mathrm{\ K}$ and $10^4\ \klvn$ as ``minihalos'', and halos with
virial temperatures above $10^4\ \klvn$ as ``large halos''.

\defcitealias{astro-ph/0403697}{FZH} \citet*{astro-ph/0403697}
\citepalias[hereafter][]{astro-ph/0403697} have created an elegant,
excursion-set-based approach to following the ionized fraction of the
Universe, which captures the clustering of ionizing sources. They
begin with the assumption that the amount of mass, $M_\mathrm{ion}$,
that sources within a given halo can ionize, is simply proportional to
the total mass of the halo,
\begin{equation}
M_\mathrm{ion} = \zeta M_\mathrm{halo} \text{,}
\end{equation}
where $\zeta$ is the ``ionization efficiency factor''. The ionized
fraction $x_\mathrm{i}$ of a region is then proportional to the
fraction of mass in that region that is contained in collapsed
objects,
\begin{equation}
x_\mathrm{i} = \zeta f_\mathrm{coll}(M, \delta, z, \Tvirmin{})
\text{,}
\end{equation}
where $M$ is the mass of the region, $\delta \equiv (\rho -
\bar{\rho})/\bar{\rho}$ is the overdensity of the region, and $z$ is
the redshift.

The collapsed fraction in the extended Press-Schechter formalism is:
\begin{equation}
  f_\mathrm{coll}(M, \delta, z, M_\mathrm{min}) = \mathrm{erfc} \left[
  \frac{\delta_c(z) - \delta(M)} {\sqrt {2[\sigma^2(M_\mathrm{min}) -
  \sigma^2(M)]}} \right] \text{.}
\end{equation}
The critical density for collapse is $\delta_c = 1.686/g(z)$, where
$g(z)$ is the linear growth factor of density perturbations,
normalized to unity at $z=0$ \citep[we use the approximation
from][]{1992ARA&A..30..499C}. The variance of $\delta(M)$,
$\sigma^2(M)$, is determined from the power spectrum, which we in turn
calculate using the fitting formulae by \citet{astro-ph/9710252}.

We adopt the relation between virial temperature and halo mass given
by \citet{2001PhR...349..125B}:
\begin{eqnarray}
T_\mathrm{vir}(M,z) &=& 1.98 \times 10^4 \mathrm{\ K\ }
\left(\frac{M}{10^8 h^{-1} \msun}\right)^{2/3}
\left(\frac{1+z}{10}\right) \nonumber \\ && \times
\left[\frac{\Omega_{M,0}}{\Omega_M(z)}\frac{\Delta_\mathrm{c}}{18
\pi^2} \right]^{1/3} \left(\frac{\mu}{0.6}\right) \text{,}
\end{eqnarray}
where $\Delta_\mathrm{c} \equiv 18 \pi^2 + 82 [1-\Omega_M(z)] - 39
[1-\Omega_M(z)]^2$, and we set $\mu = 1.22$ for $T_\mathrm{vir} < 10^4
\ \klvn$ and $\mu = 0.59$ for $T_\mathrm{vir} \ge 10^4 \mathrm{\
K}$. We solve this expression for $M$ to determine the minimum halo
mass, $M_\mathrm{min}(\Tvirmin{}, z)$, as a function of redshift.

The condition for a region of mass $M$ to be fully ionized is
$\zeta^{-1} f_\mathrm{coll} \ge 1$. \citetalias{astro-ph/0403697}
solve this equation for $\delta(M)$ yielding a barrier function
$\delta_x(M,z)$.  The largest mass at which $\delta(M) \ge
\delta_x(M,z)$ about a particular point is then identified as the mass
of the ionized region to which the point belongs.  Computing the
mass $M$ defined by this identification at different redshifts then
translates into the history for the mass of the ionized bubble in
which the given point resides.  Conversely, the mass function of
ionized bubbles at any redshift can be constructed either by Monte
Carlo realizations of many $\delta(M)$ trajectories, or from the
statistics of the first-crossing distribution of a random walk
\citep[for a method of finding the first-crossing distribution for an
arbitrarily shaped barrier, see][]{Zhang:2005ar}.

\begin{figure*}[htb]
\hbox to \hsize{\hfil\hskip-0.2in\vbox to 3.3in{
\includegraphics[height=0.45\textwidth,angle=270]{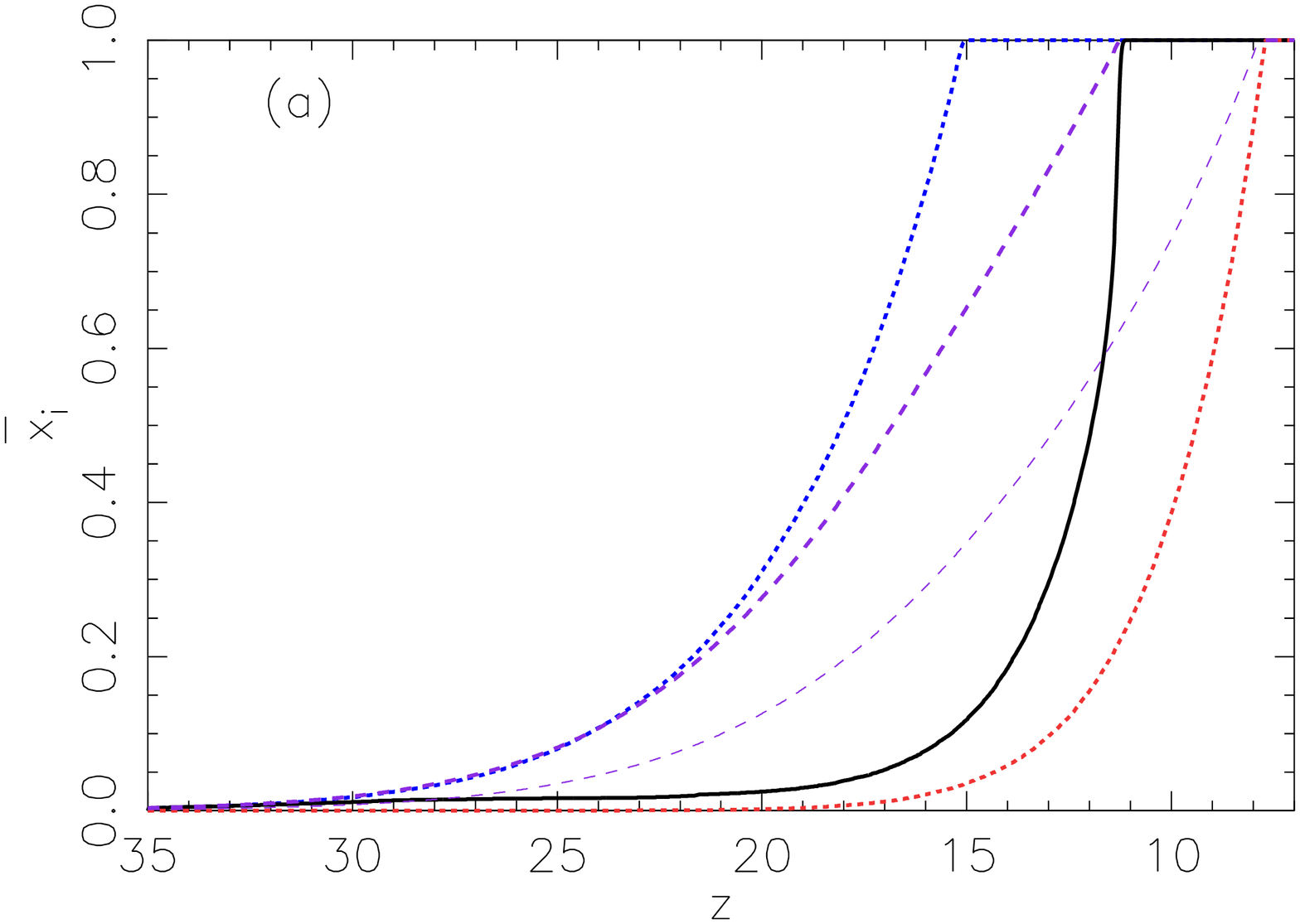}~~~
\includegraphics[height=0.45\textwidth,angle=270]{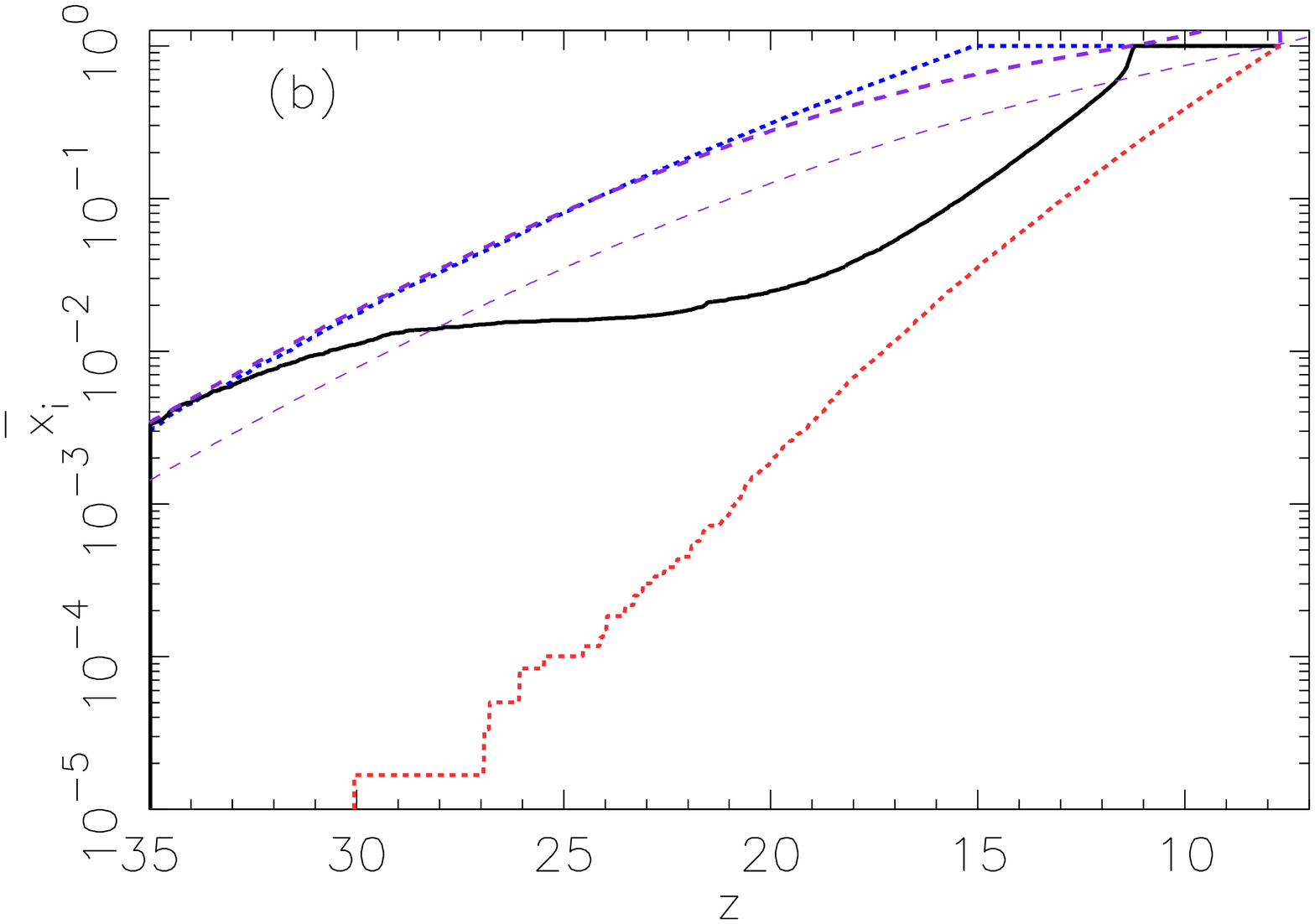}
\vfil}\hfil}
\vskip-1.in
\caption{The ionization history of the Universe: mean ionized fraction
versus redshift. The solid curve is the history with biased
feedback. The leftmost short--dashed curve (blue in the color version)
includes contributions from all halos, the rightmost short-dashed
curve (red) includes only large halos, and the upper long-dashed curve
(purple) includes unbiased feedback. All of the above use $\zeta =
12$. The light--weight long--dashed curve (purple) is an unbiased
feedback calculation with a lowered efficiency factor, $\zeta =
5$. Comparing the lower unbiased feedback curve with the biased
feedback curve demonstrates that minihalo clustering must have a large
effect on the early stages of reionization even if the reionization
redshift is assumed to be known.}
\label{reion-history}
\end{figure*}

The cosmological parameters used throughout this paper are: $h =
0.72$, $\Omega_{M,0} = 0.27$, $\Omega_{\Lambda} = 1 - \Omega_{M,0}$,
$\Omega_\mathrm{b,0} = 0.045$, $\sigma_8 = 0.9$, $n = 1$, taken from
the best--fit values in the 1--year {\it WMAP} data release
\citep{astro-ph/0302209}. For our purposes, the most significant
changes in the cosmological parameters in the 3--year {\it WMAP}
results \citep{Spergel:2006hy} are the lower values of $\sigma_8 =
0.74^{+0.05}_{-0.06}$ and $n_s=0.951^{+0.015}_{-0.019}$.  The main
effect of both of these changes is to slightly delay structure
formation and reionization.  This can be compensated for by increasing
the efficiency parameter $\zeta$, and it does not affect our primary
conclusions about the effect of clustering.

\subsection{Adding feedback}

We next depart from the \citetalias{astro-ph/0403697} formalism to
incorporate photoionization feedback. Adding feedback means that the
barrier $\delta(M)$ at any given redshift can no longer be
pre--specified; rather, it will now be a function of the entire
ionization history of the region being considered. We therefore must
resort to Monte-Carlo simulations to calculate the coupled history of
ionization and feedback around any given point.

If we assume that minihalos are randomly distributed and unbiased with
respect to the ionized bubbles, then the ionized fraction of a region
can be found by integrating the following simple differential
equation, starting from high redshift:
\begin{equation}
\frac{dx_\mathrm{i}}{dz} = 
\frac{dx_\mathrm{i,large}}{dz} + \left( 1-x_\mathrm{i}
\right)\frac{dx_\mathrm{i,mini}}{dz} \label{diff-eq}
\end{equation}
where $dx_\mathrm{i,large}/dz$ is the contribution to ionizing the
region from large halos and $dx_\mathrm{i,mini}/dz$ is the
contribution from minihalos:
\begin{align}
\frac{dx_\mathrm{i,large}}{dz} &\equiv \zeta \frac{d
  f_\mathrm{coll}}{dz}(\Tvir{} > 10^4 \ \klvn)\\
  \frac{dx_\mathrm{i,mini}}{dz} &\equiv \zeta \frac{d
  f_\mathrm{coll}}{dz}(10^2 \klvn < \Tvir{} < 10^4 \ \klvn)
\end{align}
Note that these expressions follow \citetalias{astro-ph/0403697} and
ignore recombinations (we will discuss this issue in
\S~\ref{subsection:caveats} below).  Applying these equations to the
whole Universe (i.e., to a region of mass $M \rightarrow \infty$,
$\delta = 0$) gives us the reionization history of the Universe in the
absence of any clustering.

The efficiency factor is chosen to be $\zeta = 12$, unless stated
otherwise. This factor can be thought of as the product
\begin{equation}
\zeta \equiv f_\mathrm{esc} f_* N_{\gamma/b} \left(1 +
n_\mathrm{rec}\right)^{-1}
\end{equation} 
where $f_\mathrm{esc}$ is the fraction of ionizing radiation that
escapes from halos, $f_*$ is the fraction of baryons that end up in
stars, $N_{\gamma/b}$ is the number of ionizing photons produced per
stellar baryon, and $n_\mathrm{rec}$ is the average number of times a
hydrogen atom recombines. As an example, $\zeta = 12$ could be
obtained by adopting the reasonable choices of $f_\mathrm{esc} = 0.1$,
$f_* = 0.2$, $N_{\gamma/b} = 4200$, $n_\mathrm{rec} = 6$. We ignore
the dependence of $n_{\rm rec}$ on bubble size, an approximation which
becomes inaccurate toward the tail end of reionization
\citep{astro-ph/0505065}. We discuss this issue further in
\S\ref{subsection:caveats}.

The integration is performed numerically from $z=35$ to $z=7$ with
step size $dz = 0.02$. All distances and densities specified in this
paper are in Lagrangian comoving coordinates. Lagrangian coordinates
are appropriate for our problem since we care more about the column
density between two points (which is directly related the required
number of ionizing photons) than about their physical separation.

Figure \ref{reion-history} shows the evolution of the ionized fraction
in the above toy model. The figure displays the history with only
large halos allowed to contribute (rightmost short--dashed, red,
curve), with all (i.e. both large and mini-) halos contributing
(leftmost short--dashed, blue, curve), and with minihalos suppressed
by unbiased feedback (upper long--dashed, purple, curve). The first
two curves will bound the reionization history with feedback, since
they represent the most extreme cases. As expected, the unbiased
feedback history begins following the all--halo curve, then lags
behind as the ionized fraction increases and minihalos are suppressed.

\subsection{Capturing clustering}

\begin{figure*}[htb]
\hbox to \hsize{\hfil\hskip-0.2in\vbox to 3.3in{ 
\includegraphics[width=0.45\textwidth]{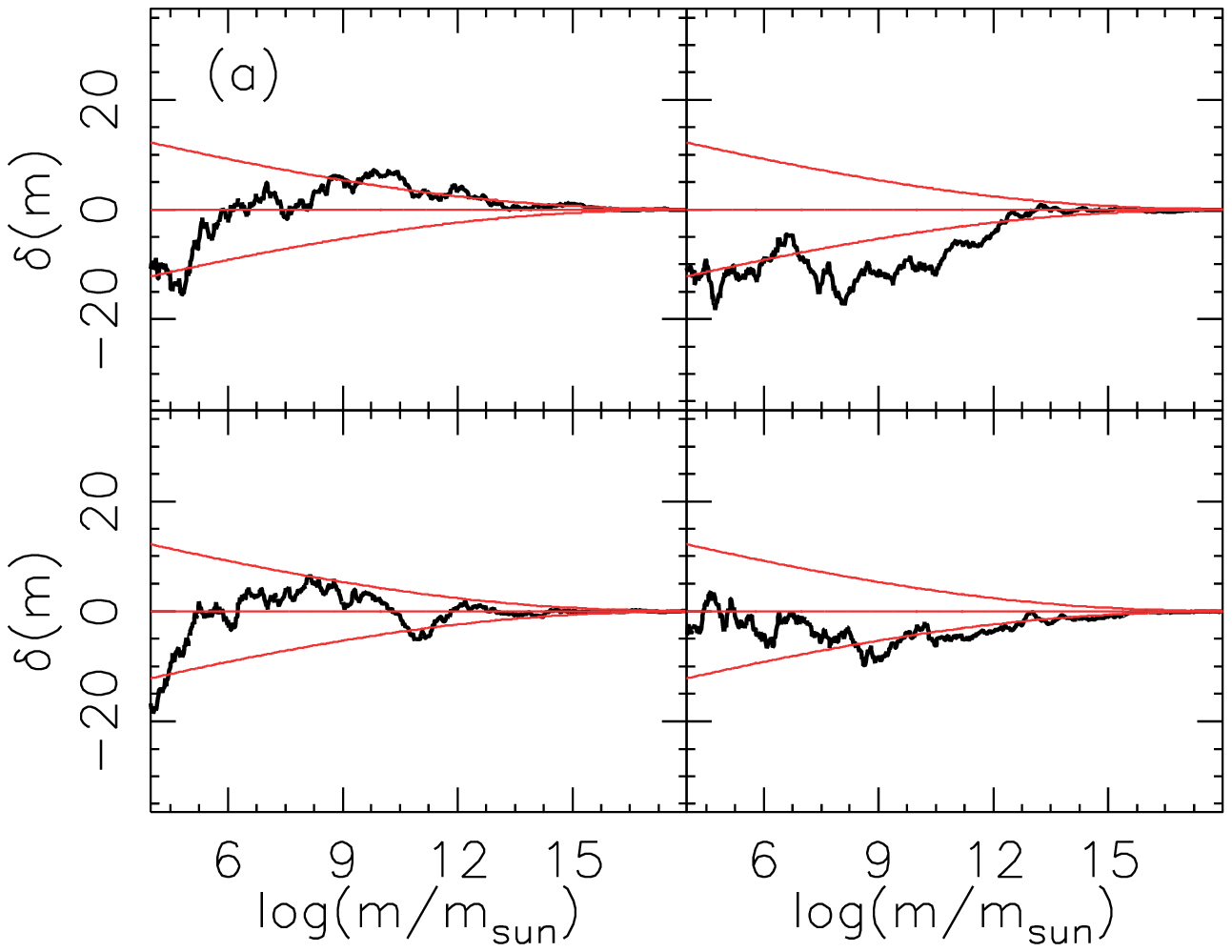}~~~
\includegraphics[width=0.45\textwidth]{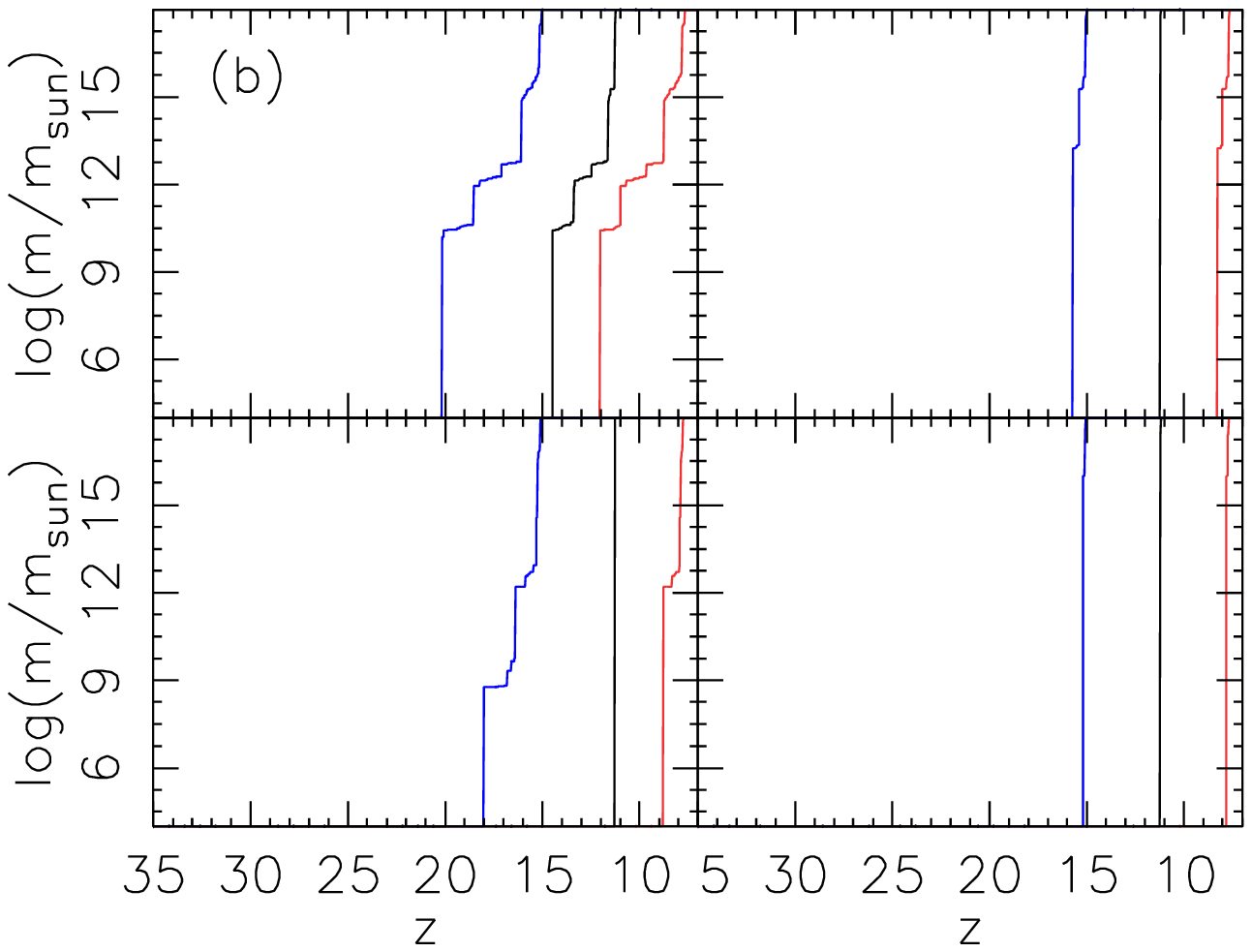}
\vfil}\hfil}
\vskip-0.8in
\caption{Examples of the density profile (a) and the corresponding HII
region mass history (b) about four randomly chosen points. Heavy
(black) lines in (a) show the mean overdensity in a spherical region
versus the mass of the region, extrapolated linearly to the
present. Light (red in the color version) curves mark the $\pm 1
\sigma$ level for the density fluctuations. In (b) the HII region mass
to which each point belongs is plotted against redshift for three
cases: (left) all (mini- and large) halos contribute to reionization,
(center) feedback is turned on, and (right) only large halos
contribute.}
\label{smallmultiples}
\end{figure*}

In order to capture the biased distribution of the minihalos with
respect to the ionized bubbles, we begin by synthesizing random walk
trajectories representing the overdensity $\delta(M)$ around a random
point in space averaged over spheres of mass $M$. Figure
\ref{smallmultiples}a shows such profiles. The overdensity tends
towards zero as $M$ gets large, since density fluctuations average out
in very large spheres.

We then divide the trajectory up into thin spherical shells and
numerically integrate equation~(\ref{diff-eq}) for each shell
separately. In this paper shells are logarithmically spaced from
$10^4$--$10^{18} \msun$ with thickness $d\log(M/\msun) =
0.015$. Shells that overproduce ionizing photons are allowed to
contribute to the ionization of the outside neighboring shell (this is
calculated recursively to allow extra photons to spill outward until
consumed). For the purpose of finding the mass of the ionized bubble
surrounding the central point, as defined above, the ionized mass is
still associated with the shell from which the ionizing photons
originated.

The total mass ionized by sources within a sphere of size $M$ is the
sum of the contribution from all interior shells,
\begin{equation}
M_\mathrm{ion}(M) = \sum_\mathrm{M' < M} M_\mathrm{ion,shell}(M') \text{.}
\end{equation}
As in the \citetalias{astro-ph/0403697} formalism, a point is said to
belong to an ionized bubble of mass $M$ such that $M$ is the largest
mass for which $M_\mathrm{ion}(M) \ge M$.

Figure \ref{smallmultiples}b shows the mass histories corresponding to
the four density profiles shown in Figure~\ref{smallmultiples}a for
the all--halo, large--halo--only, and biased feedback cases. As was
found in \citet{astro-ph/0505065}, the growth of the region proceeds
primarily by large jumps in mass, corresponding to ``major mergers'',
wherein a point is engulfed by a neighboring, larger HII region. The
mass of a merger product is determined by where a peak occurs in the
$\delta(M)$ trajectory, so many of these jumps are repeated in each
history at the same mass, but at different redshifts. This explains
why the trajectories in the three different cases in Figure
\ref{smallmultiples}b appear to be nearly identical copies of one
another, shifted in redshift.

Note that this prescription only partially captures the clustering of
minihalos. Because of the division into concentric shells, the radial
dimension of the clustering around each point is well captured, while
the clustering of halos in the tangential direction, within each thin
shell, is ignored. The model, therefore, performs best while ionized
bubbles are small; otherwise, the effects of clustering are
underestimated. This means that the predicted minihalo suppression is
a lower limit, or, equivalently, the predicted bubble masses and
ionized fractions upper limits.

The probability that a randomly selected point in space will land in
an ionized bubble of mass $M$ to $M+dM$ is equal to the fraction of
the total volume (or mass) of the IGM occupied by such bubbles,
$f(M,z) = dx_\mathrm{i}(M,z)$, where $f(M,z)$ is the fraction of
bubble mass histories falling between $M$ and $M + dM$ at redshift
$z$. The ionized fraction of the Universe is then
\begin{equation}
\bar{x}_\mathrm{i}(z) = \sum_m f(M,z) \text{.}
\end{equation}
We have performed the above calculation for an ensemble of
$N_\mathrm{total} = 60,000$ realizations of the density profile; this
implies that we can resolve a minimum global ionization fraction of
$\bar{x}_\mathrm{i} \sim 1/N_\mathrm{total} \sim 10^{-5}$.

\section{Results}
\label{section:results}
\subsection{Universal ionized fraction}\label{universal-ionized-fraction}

Figure \ref{reion-history} shows the reionization history computed
from our models.  In addition to the unbiased feedback, all--halos,
and large--halos--only cases discussed previously, the figure also
displays the result in the case with biased feedback (thick solid
black curve). Adding clustering dramatically reduces the ionized
fraction at high redshift. The biased history departs from the
unbiased history before $\bar{x}_\mathrm{i} = 0.01$ and reaches a
maximum fractional suppression relative to the unbiased case of a
factor of $11$, when $\bar{x}_\mathrm{i,bias} = 0.03$ {\it vs.}
$\bar{x}_\mathrm{i,unbias} = 0.33$ at $z=19.2$, and a maximum absolute
difference of $0.56$ when $\bar{x}_\mathrm{i,bias} = 0.22$ {\it vs.}
$\bar{x}_\mathrm{i,unbias} = 0.76$ at $z=13.6$.

The figure shows that the unbiased and biased reionization histories
(which have the same efficiency, $\zeta=12$) reach percolation at the
same redshift; i.e. they both meet at $\bar{x}_\mathrm{i} = 1$ at
$z\sim 11.1$. This is an artifact of the modeling procedure, and is not
physical. It arises because the largest, outermost shells in each
density profile will always tend to have overdensities very close to
$\delta = 0$, so their ionization histories will match that of an
unbiased Universe, and will therefore be fully ionized when the
unbiased Universe is fully ionized.  In reality, feedback would always
delay percolation, as long as the contribution of minihalos to the
ionization is not negligible ($df_{\rm coll, mini}/dz \gsim df_{\rm
coll, large}/dz$; see Fig.~\ref{minihalo-fraction-plot} below). We
also emphasize that our neglect of clustering in the tangential
direction within each shell will always increase the universal ionized
fraction over what it would be if the clustering in the tangential
direction was also included. Therefore, the difference between the
unbiased and biased histories is a lower limit, and our conclusion
that clustering significantly increases minihalo suppression is
conservative and robust.

Since observations of Gunn-Peterson troughs in quasar spectra
constrain the end of reionization to $z_\mathrm{EOR} \approx 6$, and
since the value of the efficiency parameter $\zeta$ is uncertain, a
better comparison might be between unbiased and biased trajectories
with fixed $z_\mathrm{EOR}$, rather than fixed $\zeta$. The artifact
mentioned above means that our biased feedback model becomes
unreliable near percolation, making such an exact comparison
impossible.  However, we know that our biased feedback calculation
together with the large-halo-only curve constitute upper and lower
bounds on the true ionized fraction in the biased feedback
case. Therefore, if we calculate an unbiased feedback history with an
efficiency factor lowered (we find, to $\zeta = 5$) so that it reaches
full ionization at the latest possible redshift
($z_\mathrm{EOR}\approx 7.7$, matching the large--halos--only case),
then it, together with the $\zeta = 12$ unbiased feedback curve,
constitute commensurable upper and lower bounds.

In other words, in the context of a fixed $z_\mathrm{EOR}$, the
minimum possible effect at high redshift of adding clustering is shown
by the difference between the unbiased feedback curve with $\zeta = 5$
and the biased feedback curve with $\zeta = 12$. The comparison of the
thick solid (black) and the light--weight, long--dashed (purple)
curves in Figure \ref{reion-history} clearly demonstrates that
minihalo clustering has a significant effect on the high-redshift tail
of reionization. In Figure \ref{reion-history}b, which plots the
logarithm of the ionized fraction versus redshift, it can be seen that
the biased feedback curve initially follows the all--halo curve, as
expected.  However, at $z\sim 30$, the ionized fraction reaches a
plateau, indicating that nearly all minihalo formation is suppressed
(see the next subsection); the ionized fraction does not begin to
increase again until the contribution from large halos becomes
significant.

\subsection{Ionized bubble sizes}

\begin{figure}[hbt]
\includegraphics[height=0.45\textwidth,angle=270]{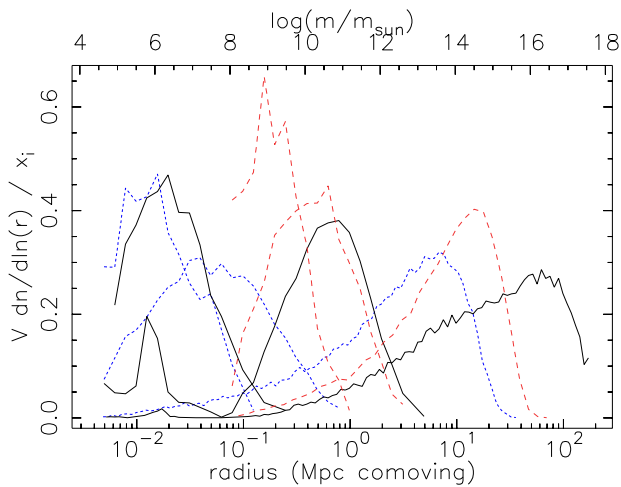}
\caption{The size distribution of HII regions shown for
$\bar{x}_\mathrm{i} = 0.01, 0.1, 0.75$ (left to right in each
set). The solid curves include biased feedback, the short--dashed
curves include all halos and the long--dashed curves are for large
halos only. The comoving bubble radius is indicated on the bottom
axis, and the equivalent mass on the top axis. The redshifts
corresponding to the ionized fractions listed above are $z=31.7, 24.3,
16.3$ with all halos, $30.7, 15.3, 11.4$ with biased feedback, and
$17.3, 12.9, 8.4$ with large halos only. \label{size-dist}}
\end{figure}

In order to better understand the results in the previous section, we
here examine the size--distribution of the ionized bubbles. The
fraction of points lying in bubbles between mass $M$ and $M + dM$ is
\begin{equation}
f(M,z) = d\bar{x}_\mathrm{i}(z) = V(M) \frac{dn_\mathrm{HII}}{dM} dM
\end{equation}
where $V(M)$ is the volume of a region of mass $M$, and
$dn_\mathrm{HII}/dM$ is the number density of ionized bubbles per unit
mass between $M$ and $M+dM$. We solve this equation to find the mass
distribution of the bubbles, given the histogram of HII region masses
derived from our ensemble of density profiles. Figure \ref{size-dist}
plots the size distribution of ionized bubbles, normalized by the
average ionized fraction of the Universe
\citepalias[after][]{astro-ph/0403697},
$V(M)~dn_\mathrm{HII}/d\ln(r)~\bar{x}_\mathrm{i}^{-1}$, for
$\bar{x}_{i}=0.01,0.1,0.75$. This function is chosen so that the area
under a segment of the curve (in log space) is equal to the fraction
of ionized gas contained in bubbles in that radius (or mass) interval.

\begin{figure*}[htb]
\hbox to \hsize{\hfil\hskip-0.2in\vbox to 3.3in{
\includegraphics[height=0.45\textwidth,angle=270]{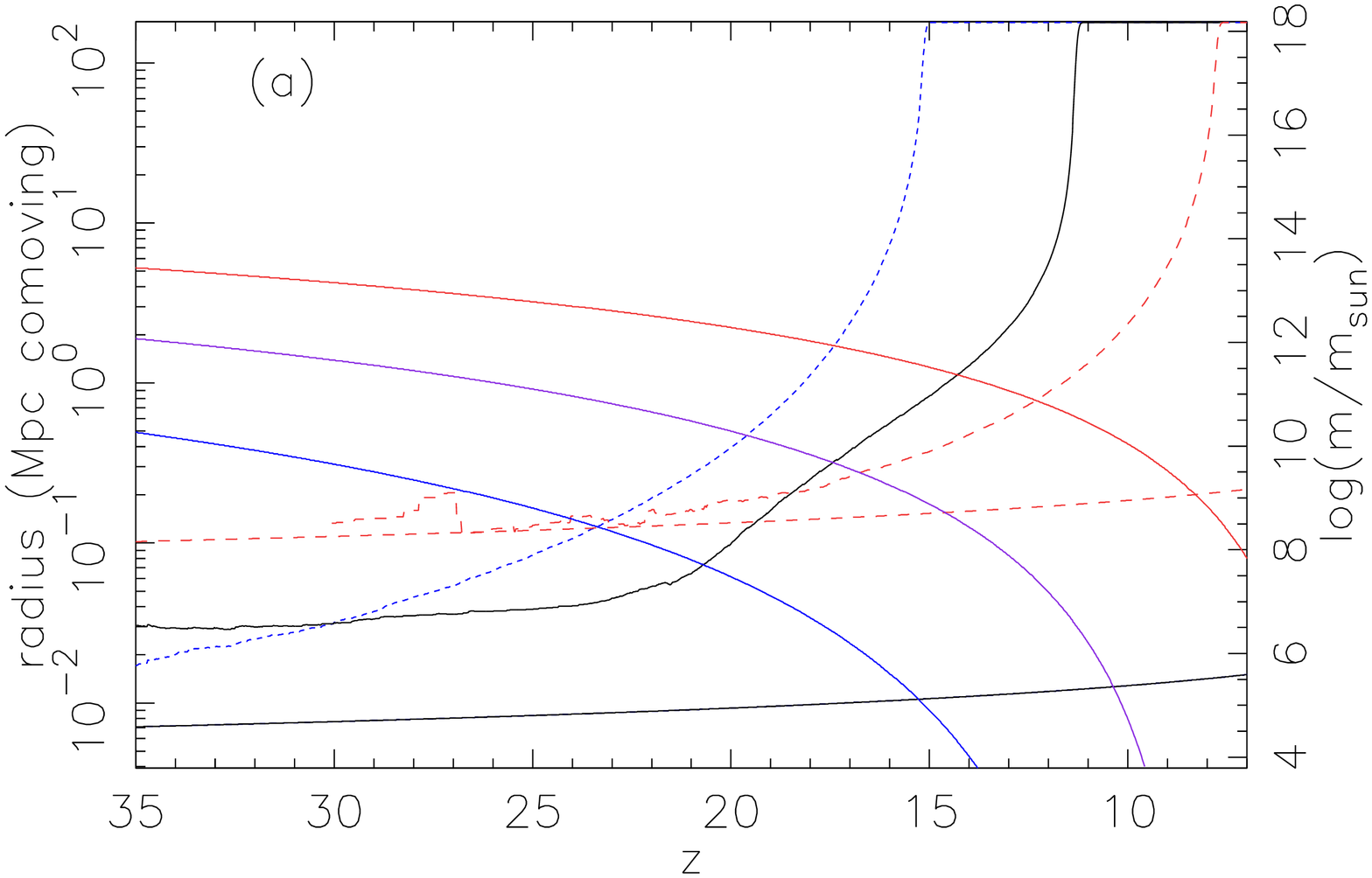}~~~
\includegraphics[height=0.45\textwidth,angle=270]{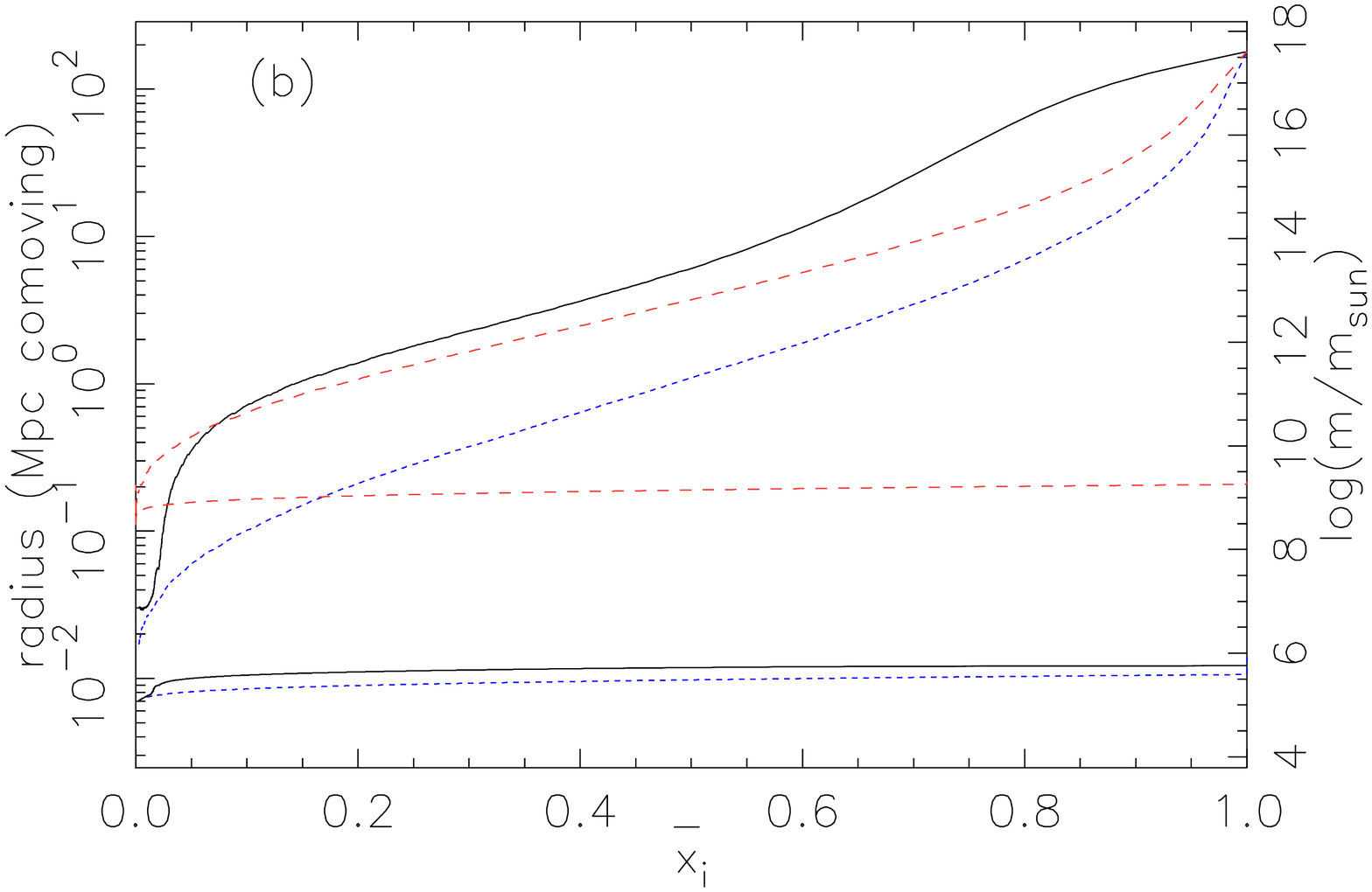}
\vfil}\hfil}
\vskip-1.2in
\caption{The mean HII region size versus redshift (a) or versus
Universal ionized fraction (b) for the large-halo-only (long--dashed),
all--halo (short--dashed), and biased feedback (solid) cases. Also
shown (nearly-flat curves) is the minimum size $M_1$ for an isolated
bubble around a minimum-mass halo for each case. The three solid
curves that curve downwards at low redshift are (from top to bottom)
the correlation lengths for large halos, the cross-correlation length
between large and minihalos, and the correlation length for small
halos.}
\label{size-mean}
\end{figure*}

The leftmost set of curves in the figure shows that when there is very
little ionized gas ($\bar{x}_\mathrm{i} = 0.01$), it is mostly
contained in bubbles around individual halos. Bubble sizes are
therefore close to $M_1 \equiv \zeta M_\mathrm{min}(z)$, the size of a
bubble around a single, minimum-mass halo. For minihalos, $M_1 \sim
10^5~\msun$, while for large halos $M_1 \sim 10^8~\msun$. The
equivalent comoving radii are $\sim 10^{-2}~\Mpc$ and $\sim
10^{-1}~\Mpc$.  

At the intermediate value $\bar{x}_\mathrm{i} = 0.1$, the feedback
case has developed a double-peaked distribution. The lower peak
corresponds to bubbles close to $M_{1} \equiv \zeta M_{\rm min}(z)$
for minihalos, while the upper peak corresponds to larger bubbles
caused by source clustering. Note that this bimodal distribution is
unique to the biased feedback scenario: since the minihalo
contribution in ionized bubbles is greatly attenuated, the number of
isolated bubbles of mass $M_{1}$ is boosted.  By the time reionization
is nearing completion, $\bar{x}_\mathrm{i} = 0.75$, the lower--radius
peak has nearly disappeared and the feedback case has a peak radius
that exceeds the other two cases.  It is easy to understand why the
bubble size in the feedback case exceeds that of the all--halos case:
the contribution of minihalos is attenuated and the larger bubble size
reflects the stronger clustering of the larger halos. Similar effects
emerge when one imposes a mass--dependent ionizing efficiency
\citep{2006MNRAS.365..115F}.  Note that, in fact, at $\bar{x}_{i} \ge
0.4$, the mean bubble size in the feedback case even exceeds that in
the large--halos--only case.  We caution, however, that this result
may be unphysical: as discussed above, our results become
quantitatively unreliable at low redshift. In particular,the mean
ionized fraction of $\bar{x}_{i} \approx 0.4$ is reached at the
redshift below which the minihalo contribution starts to increase (see
Fig.  \ref{minihalo-fraction-plot}), suggesting that our minihalo
suppression becomes only a lower limit below these redshifts. It is
not clear whether a more accurate treatment, with more minihalo
suppression, would further increase or decrease the mean bubble size.
Finally, we note that that in the biased feedback case, the
distribution of bubble sizes {\it broadens} as reionization proceeds,
whereas without feedback the bubble size distribution narrows, for
fairly general reasons \citep{2006MNRAS.365..115F}; this may be
another signature of the unphysical increase in minihalo contribution.

To gain further intuition, we explore how the mean bubble size and the
number density of bubbles evolves with redshift in each case.  We
define a characteristic bubble size as the mean mass--weighted bubble
radius:
\begin{equation}
\bar{r} = \frac{1}{\bar{x}_\mathrm{i}} 
\int r V(r) \frac{dn_\mathrm{HII}}{dr} dr \text{.}
\end{equation}
When the evolution of the characteristic size is considered in
redshift (Figure \ref{size-mean}a), $\bar{r}$ with feedback is
intermediate between the all--halo and large--halo--only cases, both
in the very beginning (at $z\gsim 30$, when isolated bubbles
dominate), and also later (at $z\lsim 18$, when enough large halos
have formed to drive the distribution). However, when $\bar{r}$ is
considered against the ionized fraction $\bar{x}_\mathrm{i}$ (Figure
\ref{size-mean}b), the radius in the feedback case quickly exceeds
that of both ``extremes'', although as noted above, the mild excess
above the large halo only case may not be physical.  Note that all
three curves should approach the Hubble radius as $\bar{x}_\mathrm{i}$
approaches unity. The unphysical convergence at $M = 10^{18} \msun$
occurs because this is the largest mass scale included in the
numerical modeling.  For reference, in Figure \ref{size-mean}, we also
show the correlation lengths, defined by $b(M_1)b(M_2)\xi(r)=1$, where
$b(M)$ is the linear bias of halos \citep{Sheth:1999mn}, and $\xi(r)$
is the correlation function of mass.

\begin{figure*}[htb]
\hbox to \hsize{\hfil\hskip-0.2in\vbox to 3.3in{
\includegraphics[height=0.45\textwidth,angle=270]{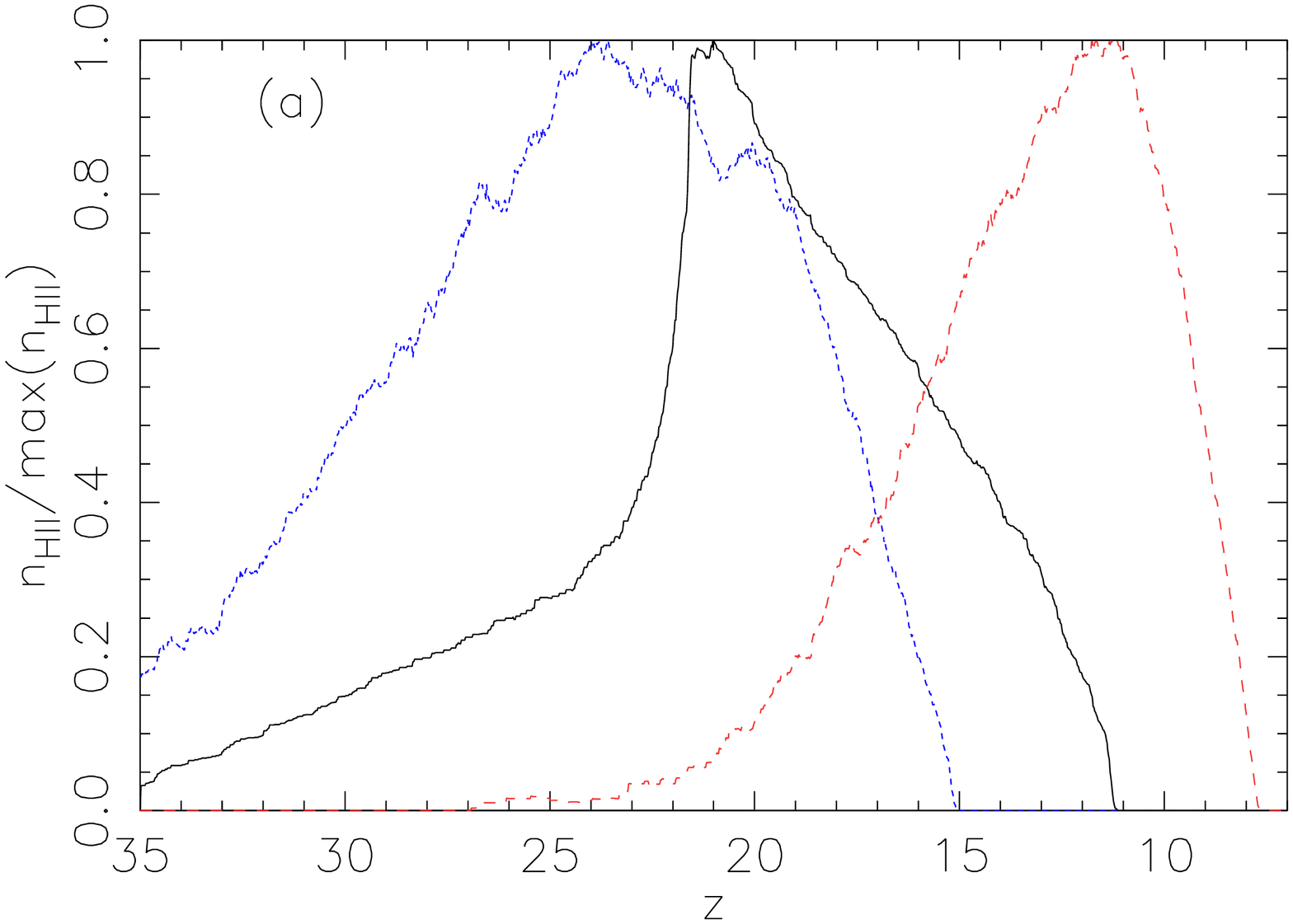}~~~
\includegraphics[height=0.45\textwidth,angle=270]{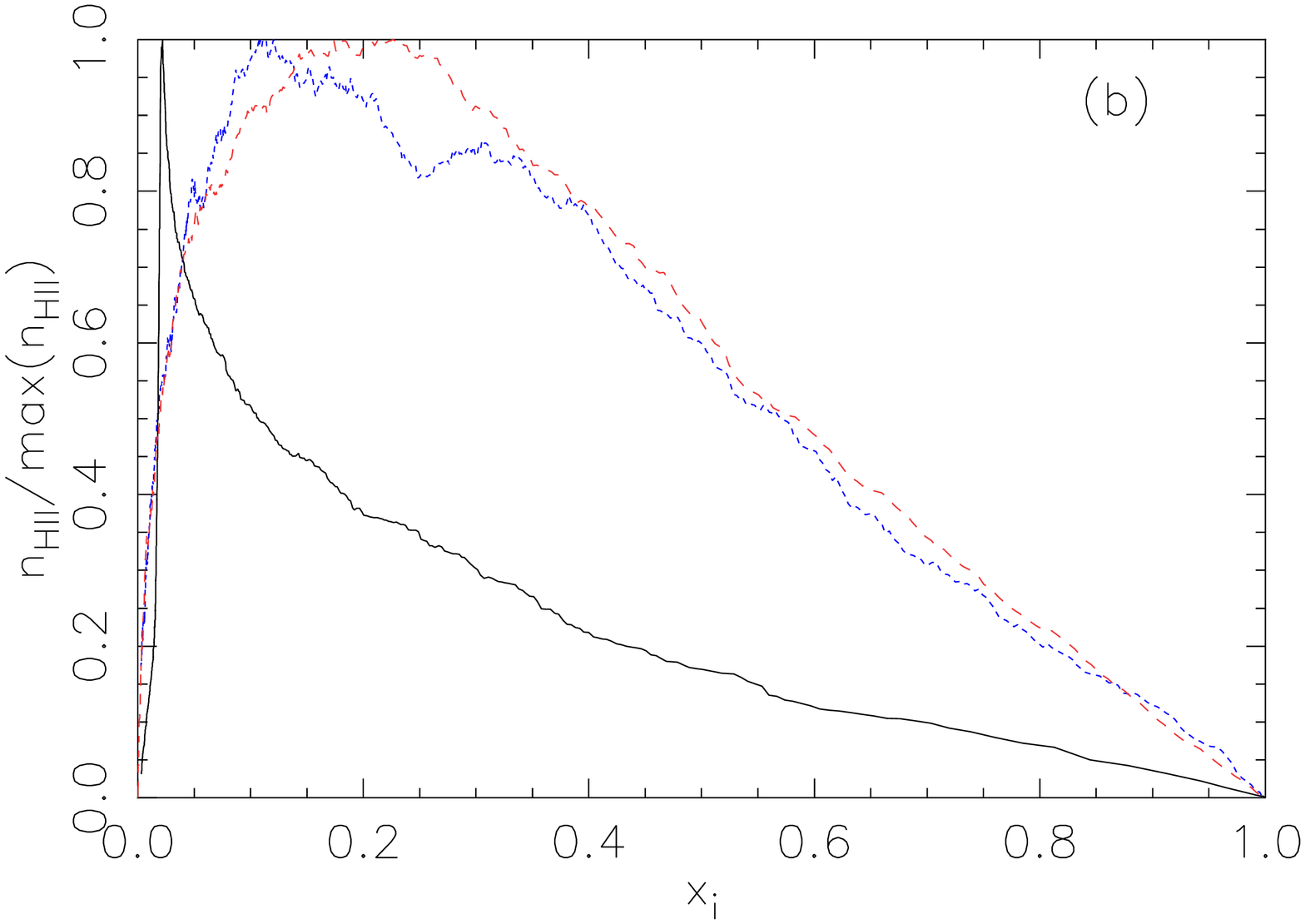}
\vfil}\hfil}
\vskip-1.in
\caption{The comoving number density of HII regions versus redshift
(a) or versus Universal ionized fraction (b) for the large--halo--only
(long--dashed), all--halo (short--dashed), and feedback (solid)
cases. The plots are normalized by the maximum number density for each
case, which is $1.0 \times 10^4~\Mpc^{-3}$ with all halos, $1.5 \times
10^4~\Mpc^{-3}$ with biased feedback, and $2.8~\Mpc^{-3}$ with large
halos only. }
\label{bubble-density}
\end{figure*}

The total number density of ionized bubbles, $n_\mathrm{HII} = \int
(dn_\mathrm{HII}/dr) dr$, is plotted versus redshift and ionized
fraction in Figure \ref{bubble-density}. Comparing this to Figure
\ref{size-mean} allows us to describe the growth of HII bubbles during
the reionization process.

During the first phase of reionization, most of the ionized gas is
contained in small bubbles around isolated halos, or groups of a few
halos. The increase in the bubble size is mostly due to the formation
of new, larger halos, and the number of HII regions climbs
steadily. This phase ends when the characteristic bubble size reaches
the correlation length of the relevant halo population. At this point,
the second phase begins. Bubbles start to merge rapidly, and the
growth of the characteristic bubble size is now primarily due to those
mergers. The number of HII regions declines steadily as the bubble
sizes grow larger and larger. Reionization is complete when the bubble
size becomes formally infinite.

In the feedback case, the transition to the second phase occurs at a
much smaller ionized fraction than in the other two cases (note the
early peak in Figure \ref{bubble-density}b). At $z\approx24$ in the
all--halo case, the bubble size exceeds the correlation length, but
this is delayed by minihalo suppression in the feedback case until
$z\approx20$. By this time there are more bubbles (even though the
ionized fraction is lower), because more isolated halos have had time
to form in slightly less-dense regions, and these bubbles have not yet
merged away.

As a result, we find that minihalo suppression slows the growth of the
bubbles and delays their mergers, while allowing more individual
bubbles to form. Because the bubbles are smaller, they can be more
tightly ``packed'', i.e. their separations are smaller, so when they
do begin to overlap, a very rapid period of bubble growth by mergers
ensues ($z\approx 22$ to $z\approx 20$). Because the mergers occur so
rapidly (and because of high minihalo suppression), the ionized
fraction grows very little, while the bubble size grows by over an
order of magnitude (note the sharpness of the peak in Figure
\ref{bubble-density}b and the rapid rise in radius in Figure
\ref{size-mean}b). This means that the mean bubble size for the
feedback case ends up very high, for any given ionized fraction at
$\bar{x}_\mathrm{i}\gsim 0.05$.

\subsection{Mean bias factor}

Our main result above can be summarized as follows: a much larger
fraction of minihalos are suppressed when their clustering is
included, relative to the value of $\bar{x}_\mathrm{i}$ expected if the
minihalos were randomly distributed.  Here we compare the ``boost'' in
the fraction of suppressed halos we find to naive expectations, using
a linear halo--bias model.

\begin{figure*}[htb]
\hbox to \hsize{\hfil\hskip-0.2in\vbox to 3.3in{ 
\includegraphics[height=0.45\textwidth,angle=270]{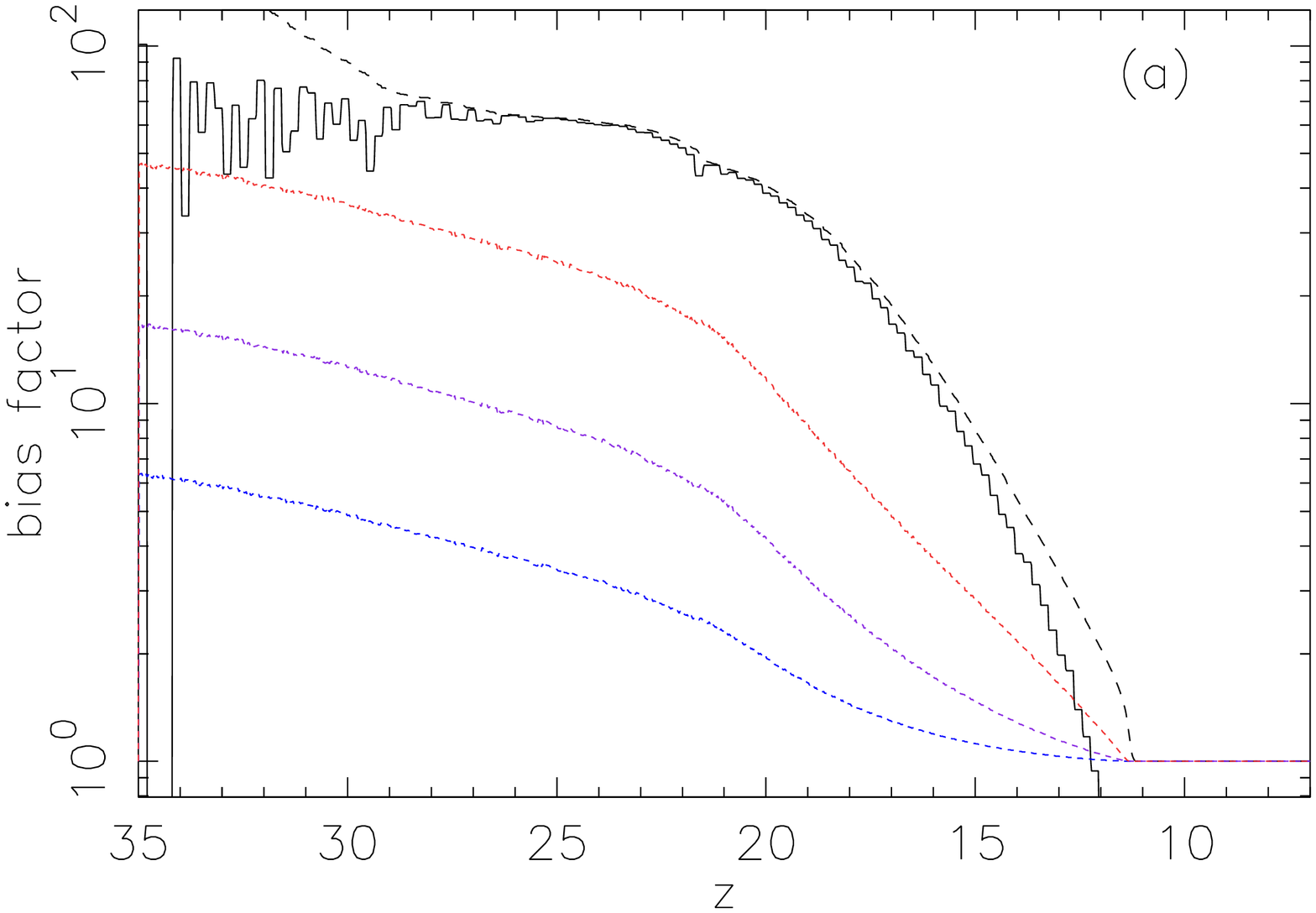}~~~
\includegraphics[height=0.45\textwidth,angle=270]{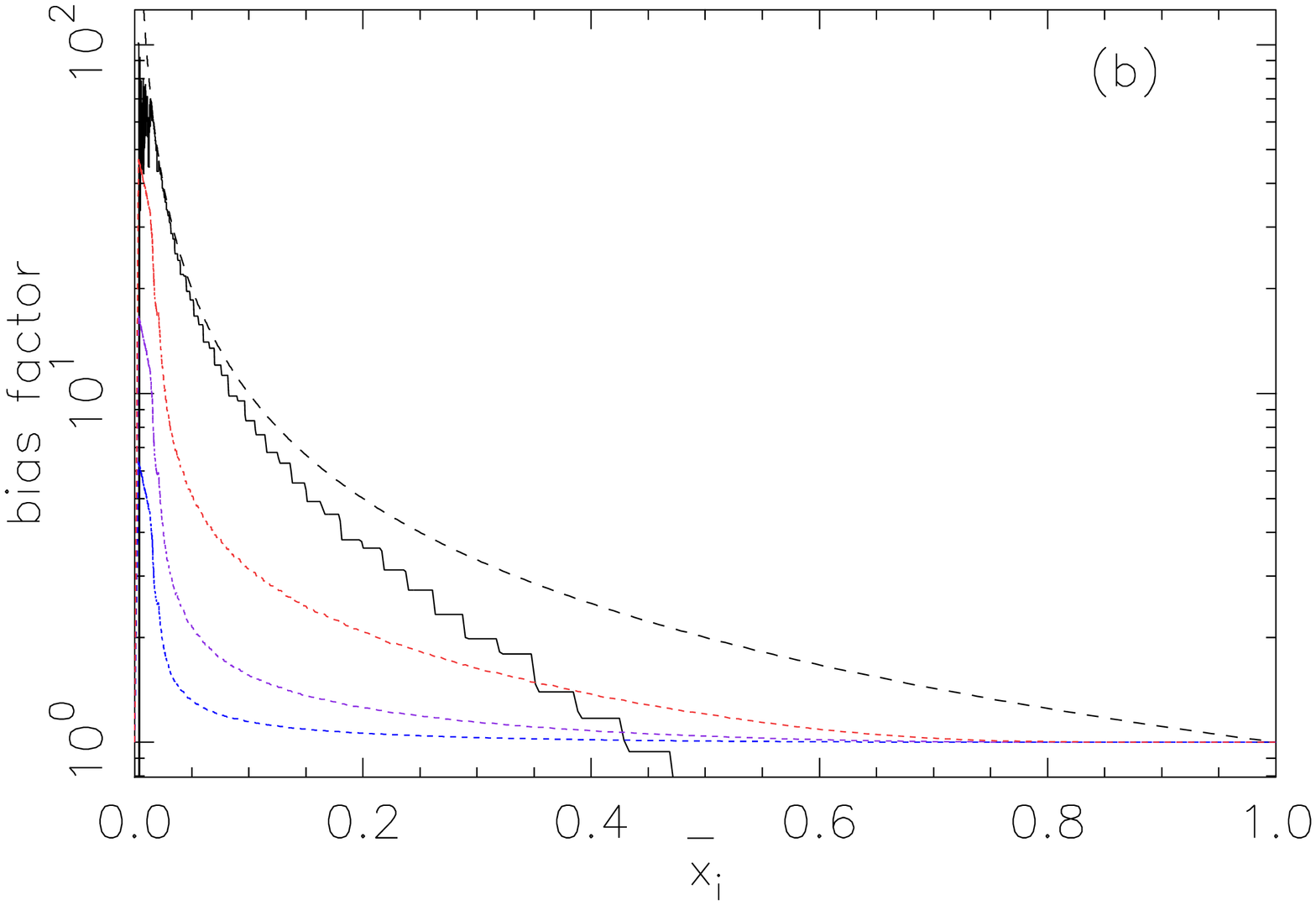}
\vfil}\hfil}
\vskip-1.2in
\caption{Mean bias enhancement factor versus redshift (a) and ionized
fraction (b), calculated using equation \ref{bias-eq} (top solid
curve). Also plotted are the enhancement in the number of neighbors
for (lower three curves, top to bottom) large halos, large with
minihalos, and minihalos, from equation \ref{bias-approx}. The
long-dashed curve shows $\bar{x}_\mathrm{i}^{-1}$ and corresponds to
the limiting case when all minihalos are suppressed.}
\label{bias-plot}
\vskip+0.4in
\end{figure*}

Let us first assume that the average ionized fraction of the Universe
evolves according to this differential equation,
\begin{equation}
\frac{d\bar{x}_\mathrm{i}}{dz} = 
\frac{d\bar{x}_\mathrm{i,large}}{dz} + \left( 1- \bar{B}(z) \bar{x}_\mathrm{i}
\right)\frac{d\bar{x}_\mathrm{i,mini}}{dz} \text{,}\label{diff-eq-bias}
\end{equation}
where $\bar{B}(z)$ is the effective ``boost'', i.e. the enhancement in
the number of minihalos forming in ionized regions, relative to
$\bar{x}_\mathrm{i}$, due to their biased distribution.  Since we know
each $d\bar{x}_\mathrm{i}/dz$ and $\bar{x}_\mathrm{i}$ from our
simulation runs, this equation serves as a definition of $\bar{B}(z)$.
Figure \ref{bias-plot} plots this factor, calculated from the biased
feedback ionization history using
\begin{equation}
\bar{B}(z) = \frac{1}{\bar{x}_\mathrm{i}}
\left( 1 - \frac{d\bar{x}_\mathrm{i}/dz - d\bar{x}_\mathrm{i,large}/dz}{d\bar{x}_\mathrm{i,mini}/dz} \right) \text{.}\label{bias-eq}
\end{equation}
Because the $d\bar{x}_\mathrm{i}/dz \ge d\bar{x}_\mathrm{i,large}/dz$,
we know that $\bar{B} \ge \bar{x}_\mathrm{i}^{-1}$. Note that $\bar{B}
= \bar{x}_\mathrm{i}^{-1}$ represents the limiting case in which all
minihalos are suppressed.

Figure \ref{bias-plot} shows that at the very beginning of
reionization, the enhancement is relatively low (much less than
$\bar{x}_\mathrm{i}^{-1}$) since few bubbles have grown large enough
to encompass more than one halo. However, the boost factor quickly
climbs to $\bar{B}(z) \approx \bar{x}_\mathrm{i}^{-1} \sim 50$, and at
$z\lsim 28$, nearly all minihalos are suppressed. This corresponds to
the plateau of the solid curve in Figure \ref{reion-history}b, and the
epoch of rapid mergers discussed in the previous subsection.  The
enhancement subsequently decreases as the ionized fraction increases
and the ionized regions are sampling regions whose density is closer
to the global average.  Values of $\bar{B}(z)<1$ are artifacts of the
model, caused by the unphysically rapid end of reionization discussed
in \S\ref{universal-ionized-fraction} above.

Naively, we might expect that the enhancement factor would be roughly
equal to the enhancement in the average number of halos within a
bubble radius of an average halo,
\begin{equation}
B_\mathrm{neighbor} =\frac{ 3 b(M_1) b(M_2) }{\bar{r}^3 -
r_\mathrm{min}^3} \int_{r_\mathrm{min}}^{\bar{r}} r^2 \xi_m(r) dr
\text{,}\label{bias-approx}
\end{equation}
where $b(M)$ is the bias of halos of mass $M$, $\xi_m(r)$ is the mass
correlation function, and $r_\mathrm{min}$ is a minimum radius chosen
to exclude overlapping halos (though the calculation is quite
insensitive to its exact value). Figure \ref{bias-plot} shows this
enhancement for large halos near large halos, large halos near
minihalos, and minihalos near minihalos (using the minimum masses of
each halo type). At the earliest stages of reionization the lowest
curve --- the enhancement in the number of minihalos within a bubble
radius of an average minihalo --- should be directly applicable. The
figure, however, shows that this underestimates the actual bias
enhancement by nearly an order of magnitude. 

Several effects may contribute to this interesting disparity.  First,
we are interested in small scales (the ionized mass is only $12$ times
the halo mass), where using the linear power spectrum, as we did in
equation~(\ref{bias-approx}), will underpredict the clustering.  As
shown by \citet{iliev_nonlinear}, the effects of non-linear bias can
boost the clustering of minihalos at high redshift on small scales
substantially over the standard linear result (for instance, at $z\sim
20$, when our typical bubble size is $\le 0.1$Mpc, Fig. 2 of
\citet{iliev_nonlinear} shows that non-linear bias can exceed the
standard linear result by an order of magnitude). They show that a
non-linear analytic technique based on extended Press-Schechter theory
provides an accurate match to numerical simulations; our excursion-set
based formalism, although different in detail, should likewise capture
non-linear bias. Such effects could in principle be largely
responsible for the difference between our results and
equation~(\ref{bias-approx}), although the overall contribution is
difficult to assess since volume--exclusion effects will reduce
clustering on small scales and will compensate somewhat for our
neglecting nonlinear effects. It clearly dominates at the beginning of
reionization, when most bubbles have a typical size $M_{1} \equiv
\zeta M_{\rm min}(z)$. Second, equation~(\ref{bias-approx}) gives the
increase in the number of neighbors centered around a {\it single}
halo, whereas in our case, the ionized region is already known to
contain {\it several} halos.  Third, equation~(\ref{bias-approx})
averages the bias over all halos at a given redshift; in our case, we
are interested in the number of new halos forming around a set of
halos that had formed earlier.  Fourth, the bubble-size distribution
in the feedback case always has a tail extending to small sizes, so
using the mass weighted mean radius $\bar{r}$ will underestimate the
bias; in principle, one should integrate over the distribution of
bubble sizes.

In summary, we conclude that given all the caveats and complications
mentioned above, either a tailored semi-analytic approach or a naive
estimate based on linear theory will both give misleading
underestimates of the impact of clustering.

\subsection{Minihalo suppression and electron scattering optical depth}

\begin{figure}[htb]
\includegraphics[height=0.45\textwidth,angle=270]{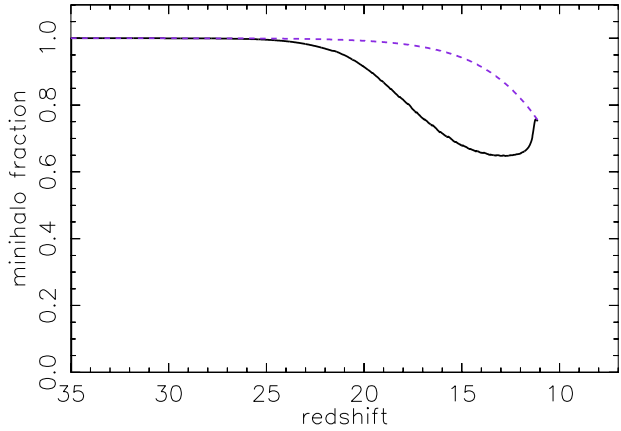}
\caption{Fraction of reionization due to minihalos for the unbiased
  (dashed) and biased (solid) feedback
  scenarios. \label{minihalo-fraction-plot}}
\end{figure}

In Figure \ref{minihalo-fraction-plot}, we show the fraction of the
reionization contributed by minihalos as a function of redshift in the
biased and unbiased feedback cases. For the reasons discussed
previously, the minihalo fraction calculated here for the biased
feedback case is an upper limit. In particular, the apparent upturn of
the minihalo fraction at low redshift is unphysical. The most
conservative conclusion is that less than $60\%$ of the full
ionization would be contributed by minihalos in a more exact treatment
of clustered feedback. It is likely, however, that in the absence of
the unphysical upturn, the minihalo contribution would fall
significantly below this value at $z\lsim 15$.

Finally, in light of the recent measurement in the three--year {\it
WMAP} data, it is interesting to compute the impact of minihalo
suppression on the optical depth $\tau$ to electron scattering.  We
compute the value of $\tau$ using the evolution of the ionized
fraction of hydrogen in our models. We ignore additional electrons
from ionized Helium (if the fraction of singly ionized helium would
track that of hydrogen, $\tau$ would increase by $\sim 8\%$).

\begin{deluxetable}{rrrrrr}
 \tablecaption{Optical depth and end-of-reionization redshift for each
   scenario\label{optical-depth-table}}
\startdata

 \tablehead{\colhead{Scenario:} & \colhead{all} & \colhead{unbiased} &
 \colhead{unbiased} & \colhead{clustered} & \colhead{large only}}\\
$\zeta$ & 12 & 12 & 5 & 12 & 12 \\
$\tau$ & 0.21 & 0.20 & 0.14 & 0.12 & 0.08\\
$z_\mathrm{EOR}$ & 15.1 & 11.1 & 7.8 & 11.1 & 7.7\\

\enddata

\end{deluxetable}

The optical depths for each history are reported in Table
\ref{optical-depth-table}. We naturally find the earliest ionization
in the no feedback (all--halos) case, so this limiting model has the
highest optical depth. With unbiased feedback, reionization is
delayed, and we find a slightly lower optical depth. The lower
efficiency unbiased case has even lower optical depth. And our history
with clustered feedback results in lower optical depth than any other
scenario, except the case in which we eliminate all minihalos.  Note
that since the biased feedback case is an upper limit on the ionized
fraction and reaches $\bar{x}_\mathrm{i} = 1$ artificially early, the
calculated optical depth is also an upper limit in this case. On the
other hand, the optical depth in the large-halo-only case constitutes
a lower limit to the true biased feedback case.

In conclusion, Table~\ref{optical-depth-table} shows that clustering
reduces the optical depth from somewhere in the range of $0.14 < \tau
< 0.20$ to somewhere in the range of $0.08 < \tau < 0.12$.

\section{Discussion}
\label{section:discussion}

\subsection{Caveats}
\label{subsection:caveats}

Obviously many simplifications must be made to construct a
semi--analytical model of a process as complicated as reionization. We
have already discussed the fact that our results for feedback with
clustering constitute an upper limit on the ionized fraction due to
the averaging of the suppression effect over spherical shells and the
dominance of large shells when the bubbles are large. However, we only
start to see obviously unphysical effects ($\bar{B} < 1$, or an
increasing minihalo contribution to reionization) at
$\bar{x}_\mathrm{i} \gtrsim 0.4$ or $z \lesssim 12$.  We expect that a
more exact treatment that incorporates feedback with a
three--dimensional treatment of clustering would stay close to this
upper limit prior to this redshift.

Another important oversimplification is that we have ignored
recombinations (except in a very rough way in the $\zeta$ factor). In
particular, we ignore the scale dependence of the recombination rate;
as the bubbles get bigger and the mean free path of ionizing photons
increases, the gas in the IGM is likely ionized up to larger
overdensities, and the importance of recombinations will increase.
The omission of recombination may also be important in the
``stalling'' regime we find occurs at high redshift $z\gsim 20$: since
there are few ionizing photons being produced and the ionized bubbles
are in relatively dense regions, recombinations would likely lower the
ionized fraction.  \citet{astro-ph/0505065} describe a method for
including recombinations in the \citetalias{astro-ph/0403697}
formalism, which might be adapted for use in this model. Although
their results suggest that recombinations are only important in the
very late stages of reionization, the larger bubble sizes produced in
the feedback case may make recombinations important somewhat
earlier. However, since the late-time behavior of our biased feedback
model when $\bar{x}_{i} \rightarrow 1$ is in any case unreliable (as
discussed in \S\ref{section:model}), we have foregone this extra
complication in the present paper.

\citet{2006MNRAS.365..115F} and \citet{CC2006} have explored several
generalizations of the \citetalias{astro-ph/0403697} formalism,
including the use of a more suitable halo mass function, stochastic
fluctuations in the halo distribution, a mass-dependent efficiency
factor $\zeta(M)$, and the history of halo mergers within ionized
bubbles. Their results suggest that the first two modifications would
have relatively minor effects on our results. A mass-dependent
efficiency, or associating sources with halo mergers, however, could
both change the importance of minihalo suppression, since it could
change the relative contributions of minihalos and large halos to
reionization. In particular, if the efficiency of ionizing photon
production in minihalos were reduced enough (by internal feedback, for
instance), minihalo suppression in ionized regions would become
unimportant.

Our specific assumptions about minihalo suppression are also somewhat
idealized. For instance, there will clearly not be a sharp cut-off of
minihalo formation at precisely $10^4\ \klvn$. There is also some
uncertainty about the exact value of the minimum virial temperature,
which may have some environmental dependence, in any case. Changing
the dividing line between minihalos and large halos will change their
relative populations and affect the precise reionization history, but
small changes should not affect the importance of minihalo suppression
as a feedback mechanism in general.

\subsection{Possible extensions}

There are several extensions to the formalism presented here that
might be profitably explored in the future, such as including a
mass-dependent efficiency factor, $\zeta(M)$, and in particular using
different efficiencies for large and minihalos. There are
possibilities for improving the performance of the model at high
$\bar{x}_\mathrm{i}$, which would allow us to better constrain the
end-of-reionization redshift and the optical depth, or, more
importantly, to calculate $\tau$ under various assumptions given a
fixed $z_\mathrm{EOR}$. As mentioned above, our formalism can be
easily adapted to treat a variety of feedback mechanisms other than
minihalo suppression; our results suggest that the impact of these
other feedback mechanisms will generically be increased by the
presence of clustering, as well. We plan to apply the formalism
discussed here to address the enrichment of the IGM with metals in a
future paper.

\section{Conclusions}
\label{section:conclusions}

Using a unique semi-analytical technique, we have demonstrated that
minihalo suppression in ionized regions is enhanced by clustering, and
that such enhanced suppression substantially reduces the high-redshift
tail of reionization. Other feedback mechanisms may be similarly
enhanced by clustering, since feedback process tend to operates over a
limited distance, especially when sources are young.

The compressed ionization history produced by minihalo suppression
reduces the optical depth $\tau$ compared to cases without clustering
or without feedback. The reduction in $\tau$ occurs both with fixed
ionization efficiency factor $\zeta$ and with fixed
end-of-reionization redshift $z_\mathrm{EOR}$.

\citet{Haiman:2006si} have shown that substantial minihalo suppression
is necessary to simultaneously meet the constraints imposed by
Gunn-Peterson troughs on $z_\mathrm{EOR}$ and by the value of $\tau$
measured in the three--year data of the {\it WMAP} experiment. We
conclude that a simple feedback mechanism --- such as preventing the
formation of minihalos in ionized regions --- can produce such
suppression when the highly-clustered distribution of these sources is
taken into account.

\acknowledgments{ZH acknowledges partial support by NASA through
grants NNG04GI88G and NNG05GF14G, by the NSF through grants
AST-0307291 and AST-0307200, and by the Hungarian Ministry of
Education through a Gy\"orgy B\'ek\'esy Fellowship. SPO acknowledges
NSF grant AST 0407084 and NASA grant 05-ATP05-115 for support.}

\bibliography{ms}

\end{document}